\documentclass[10pt,twocolumn]{article}

\usepackage[utf8]{inputenc}
\usepackage[T1]{fontenc}
\usepackage{amsmath,amssymb,amsfonts}
\usepackage{graphicx}
\usepackage[margin=0.75in]{geometry}
\usepackage{hyperref}
\usepackage{url}
\usepackage{booktabs}
\usepackage{caption}
\usepackage{subcaption}
\usepackage{natbib}
\usepackage{microtype}

\title{Synthetic Data Augmentation for Medical Audio Classification: A Preliminary Evaluation}

\author{
  David McShannon$^{1}$, Anthony Mella$^{1}$, Nicholas Dietrich$^{2}$\\
  $^{1}$Independent Researcher\\
  $^{2}$University of Toronto
}

\date{}

\begin{document}

\maketitle

\begin{abstract}
Medical audio classification remains challenging due to low signal-to-noise ratios, subtle discriminative features, and substantial intra-class variability, often compounded by class imbalance and limited training data. Synthetic data augmentation has been proposed as a potential strategy to mitigate these constraints; however, prior studies report inconsistent methodological approaches and mixed empirical results. In this preliminary study, we explore the impact of synthetic augmentation on respiratory sound classification using a baseline deep convolutional neural network trained on a moderately imbalanced dataset (73\%:27\%). Three generative augmentation strategies (variational autoencoders, generative adversarial networks, and diffusion models) were assessed under controlled experimental conditions. The baseline model without augmentation achieved an F1-score of 0.645. Across individual augmentation strategies, performance gains were not observed, with several configurations demonstrating neutral or degraded classification performance. Only an ensemble of augmented models yielded a modest improvement in F1-score (0.664). These findings suggest that, for medical audio classification, synthetic augmentation may not consistently enhance performance when applied to a standard CNN classifier. Future work should focus on delineating task-specific data characteristics, model–augmentation compatibility, and evaluation frameworks necessary for synthetic augmentation to be effective in medical audio applications.
\end{abstract}

\section{Introduction}

Medical audio classification is challenging due to low signal-to-noise ratios, subtle discriminative acoustic features, and substantial intra-class variability \cite{nanni2021ensemble,xiao2024lungadapter,dietrich2026evaluating}. These difficulties are further compounded by class imbalance and limited availability of labeled data \cite{dablain2022understanding}. In respiratory sound analysis, disease-positive recordings are particularly scarce, reflecting disease prevalence, practical barriers to clinical data collection, and privacy constraints \cite{yu2025advances,rocha2021automatic,rocha2019open}. As a result, models trained on such datasets may demonstrate reduced performance for minority classes, which often correspond to clinically important disease states \cite{rocha2021automatic,rocha2019open}.

Synthetic data augmentation using deep generative models has been proposed as a potential approach to address data scarcity and class imbalance by generating additional disease-positive samples \cite{pezoulas2024synthetic,zhu2026audio,waseem2025review,dietrich2025greener}. Recent advances in generative modeling, including Variational Autoencoders (VAEs), Generative Adversarial Networks (GANs), and diffusion-based models, have achieved strong performance in natural image synthesis and augmentation tasks \cite{gonzales2023synthetic,pasculli2025synthetic}. These developments have prompted growing interest in extending synthetic augmentation techniques to medical audio data \cite{nanni2021ensemble,xiao2024lungadapter,rocha2021automatic,bae2024patch,kala2023constrained}.

Despite this interest, generating high-quality synthetic medical audio remains difficult. The fidelity of synthetic data is constrained by the quality and diversity of available source recordings, which are often limited and noisy in clinical settings \cite{waseem2025review,pasculli2025synthetic,kala2023constrained,ghosh2025synthio}. Consequently, the practical value of synthetic augmentation for medical audio classification under realistic low-data conditions remains uncertain.

In this preliminary study, we explore the impact of synthetic data augmentation on medical audio classification performance. We evaluate commonly used generative augmentation approaches and assess their effects on downstream classification using a deep convolutional neural network (CNN) trained from random initialization.

\section{Methods}

\subsection{Dataset}

A publicly available COVID-19 dataset (Project Coswara \cite{bhattacharya2023coswara}) was selected to represent a moderate imbalance scenario. This dataset contains self-recorded respiratory sounds collected via a web application, including cough, breathing, and speech samples. For this study, we utilized cough recordings from participants labeled as either healthy or infected (confirmed COVID-19 positive), excluding those marked as exposed or recovered. This yielded 4,963 cough samples: 3,847 from healthy participants and 1,116 from COVID-19 infected participants, representing an approximate 3.4:1 class ratio. All recordings were resampled to 16 kHz mono and segmented to fixed 3-second duration using center-cropping for longer samples and zero-padding for shorter samples. The dataset was partitioned using stratified random sampling into training (70\%), validation (15\%), and test (15\%) sets, maintaining class proportions across splits. Synthetic augmentation was applied only to the training set; validation and test sets contained exclusively real recordings.

\subsection{Data Preprocessing}

Audio preprocessing followed standard practices for acoustic classification tasks. Raw waveforms were converted to mel-spectrograms using a Short-Time Fourier Transform with 2048-point FFT, 512-sample hop length, and Hann windowing. Mel-frequency representations used 128 mel-bins spanning 0--8000 Hz. Spectrograms were normalized using per-channel z-score normalization computed on the training set.

\subsection{Baseline CNN Model}

A CNN architecture designed for spectrogram classification, consisting of four convolutional blocks (32, 64, 128, 256 filters, each $3 \times 3$ kernel) with batch normalization, ReLU activation, and $2 \times 2$ max pooling. The model was trained for 100 epochs using cross-entropy loss, Adam optimizer (lr=0.001 with cosine annealing), and early stopping based on validation macro F1 score (patience=15 epochs). The classifiers were trained from scratch to isolate the effect of synthetic augmentation from potential benefits of external knowledge.

\subsection{Synthetic Data Generation}

\textbf{Variational Autoencoders:} VAEs were trained to learn a continuous latent representation of mel-spectrograms. The encoder consisted of four convolutional blocks (each with two Conv2D layers with 64, 128, 256, and 512 filters respectively, followed by batch normalization and LeakyReLU activation), followed by global average pooling and fully connected layers projecting to a 128-dimensional latent space. The decoder mirrored this architecture using transposed convolutions. The model was trained with a composite loss combining mean squared error reconstruction loss and KL-divergence regularization ($\beta = 0.1$) for 200 epochs using Adam optimizer (lr=0.0001). Synthetic samples were generated by sampling from the latent prior $\mathcal{N}(0,I)$ and decoding.

\textbf{Generative Adversarial Networks:} We employed a Wasserstein GAN with gradient penalty (WGAN-GP) to address training instability. The generator architecture consisted of dense layers expanding a 128-dimensional noise vector, followed by four transposed convolutional blocks with progressive upsampling. The discriminator used four convolutional blocks with spectral normalization. Both networks used batch normalization and LeakyReLU activations. Training proceeded for 300 epochs with critic-to-generator update ratio of 5:1, using RMSprop optimizer (lr=0.00005) and gradient penalty coefficient $\lambda = 10$. Samples were generated from Gaussian noise $z \sim \mathcal{N}(0,I)$.

\textbf{Diffusion Models:} We implemented a U-Net-based \cite{ronneberger2015unet} denoising diffusion probabilistic model. The U-Net architecture featured five encoder and decoder blocks with skip connections, incorporating multi-head self-attention at the $64 \times 64$ resolution bottleneck. The forward diffusion process added Gaussian noise over $T = 1000$ timesteps following a linear variance schedule from $\beta_{1} = 0.0001$ to $\beta_{T} = 0.02$. The model was trained to predict the noise component using a simplified L2 loss for 400 epochs with AdamW optimizer (lr=0.0001, weight decay=0.01). During inference, we used Denoising Diffusion Implicit Model sampling with 50 steps for faster generation while maintaining sample quality.

All generative models were trained exclusively on minority class (COVID-positive) samples. For each training run, synthetic samples were generated to augment the minority class by 50\% (i.e., adding 558 synthetic samples to the 1,116 real COVID-positive samples).

\subsection{Ensemble Methods}

An ensemble approach was used to combine predictions from multiple models trained under different augmentation conditions. At inference time, each of the four models: the baseline model and the models trained with VAE-, GAN-, and diffusion-based augmentation, independently processed the input mel-spectrogram and produced class probability estimates.

For each test sample, the predicted probability of the COVID-positive class from model $i$, denoted $p_{i}(\text{COVID}+)$, was averaged across models to obtain the ensemble prediction:
\begin{equation}
P_{\text{ensemble}}(\text{COVID}+) = \frac{1}{4}\sum_{i=1}^{4}p_{i}(\text{COVID}+)
\end{equation}

The final class label was assigned based on the class with the highest averaged probability. This ensemble strategy required no additional training beyond the individual models and assigned equal weight to each constituent model.

\subsection{Evaluation Protocol}

Classification performance was evaluated using macro-averaged F1 score and area under the receiver operating characteristic curve (AUROC). Macro-averaged F1 score was used as the primary metric because it equally weights performance across classes and is appropriate for imbalanced classification settings. AUROC was computed by varying the classification threshold applied to predicted probabilities and summarizes model discrimination performance across all possible thresholds. Both metrics were computed on the held-out test set, which contained only real (non-synthetic) samples.

\section{Results}

The baseline convolutional neural network achieved a macro-averaged F1 score of 0.645. VAE-based augmentation resulted in an F1 score of 0.646 (+0.001). GAN-based augmentation resulted in an F1 score of 0.609 ($-$0.036). Diffusion-based augmentation resulted in an F1 score of 0.644 ($-$0.001). The ensemble approach resulted in an F1 score of 0.664 (+0.019).

\begin{figure}[h!]
    \centering
    \includegraphics[width=1\linewidth]{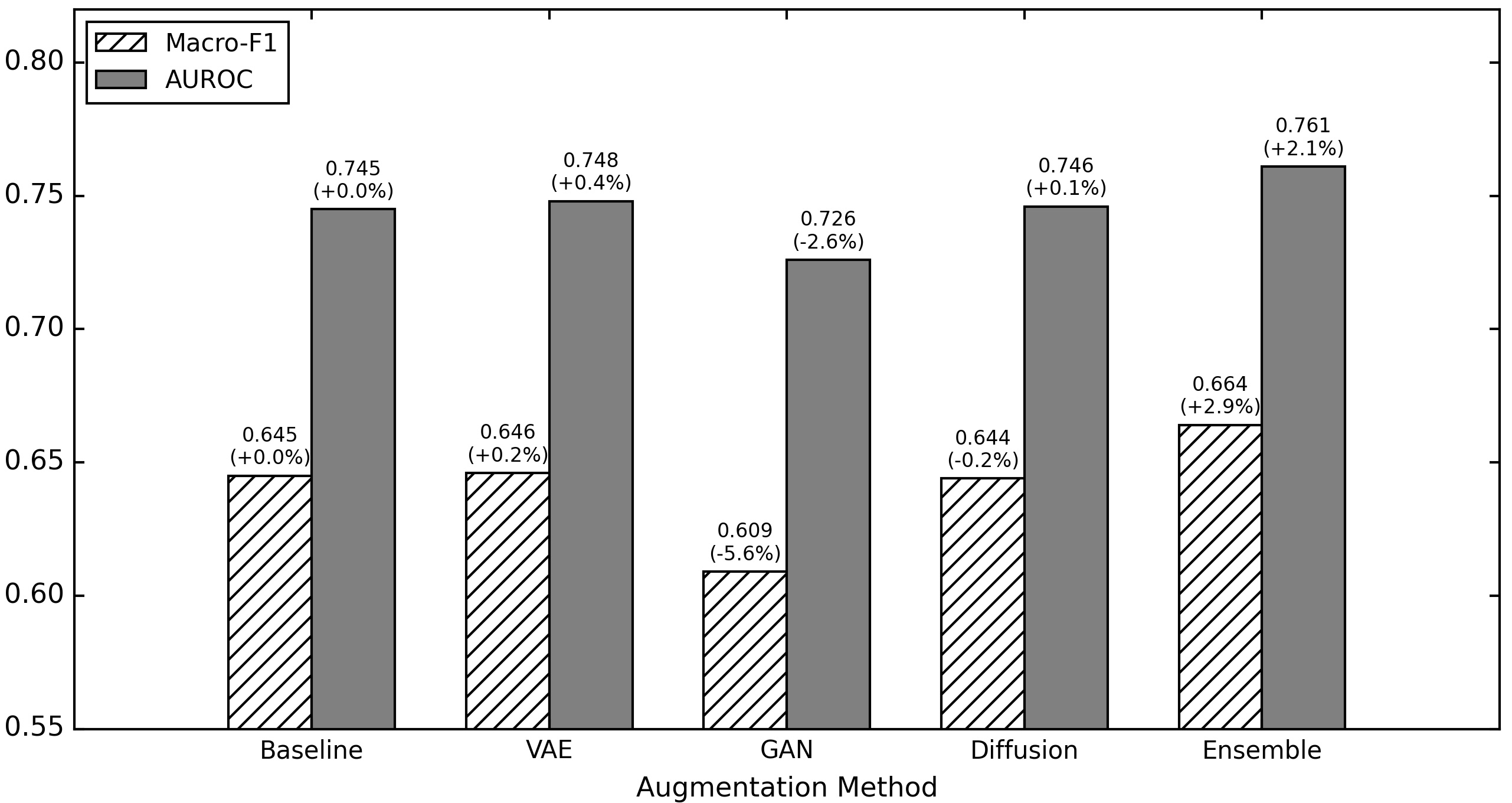}
\caption{Macro-averaged F1 score and AUROC for CNN classifiers across augmentation methods.}
\label{fig:results}
\end{figure}

AUROC values followed a similar distribution. The baseline CNN achieved an AUROC of 0.745. VAE-based augmentation resulted in an AUROC of 0.748, diffusion-based augmentation achieved 0.746, and GAN-based augmentation produced a lower AUROC of 0.726. The ensemble model achieved the highest AUROC (0.761). Figure~\ref{fig:results} shows macro-averaged F1 score and AUROC across augmentation methods.

\section{Discussion}

This preliminary study evaluated whether synthetic data augmentation improves respiratory sound classification under a moderately imbalanced, low-data setting. Across three commonly used generative approaches, we observed no consistent improvement in classification performance when synthetic samples were directly incorporated into the training set. These findings suggest that, in this setting, synthetic data augmentation may not reliably enhance performance for medical audio classification using a standard convolutional neural network.

Our findings are broadly consistent with prior work in medical audio classification \cite{dietrich2026evaluating,yu2025advances,rocha2021automatic,rocha2019open,prodhan2008wheeze,kim2021respiratory}, where gains from synthetic augmentation have been inconsistent and task dependent. Several factors may explain the limited benefit observed with individual augmentation strategies. First, medical audio signals are characterized by subtle and overlapping acoustic features, particularly for respiratory sounds such as coughs, where pathological and healthy samples share similar temporal and spectral structure \cite{rocha2021automatic,sarkar2015auscultation}. Generative models trained on limited and noisy minority-class data may struggle to learn discriminative variations, instead reproducing dominant patterns or noise present in the source recordings \cite{dablain2022understanding,chu2023cnn}. This limitation is likely amplified when the minority class contains relatively few unique examples, restricting the diversity available for generative learning.

Interestingly, the ensemble approach yielded a modest improvement in performance. This improvement emerged despite that most individual augmented models underperformed compared to baseline. This finding suggests that different augmentation strategies may induce distinct and partially uncorrelated error patterns, which can be exploited through ensembling \cite{nanni2021ensemble,karim2025ensemble,nadkarni2024afen}. In this context, the value of synthetic augmentation appears to lie less in improving individual model training and more in promoting diversity across models. However, this benefit comes at the cost of increased computational complexity and training time, which may limit the practicality of such approaches in resource-constrained clinical settings \cite{dietrich2025greener}.

Future work should explore alternative roles for synthetic data beyond direct training set augmentation \cite{xiao2024lungadapter,dietrich2025agentic,qureshi2026improving,chakraborty2025generative}. Synthetic samples may be more effective for pretraining, representation learning, or self-supervised objectives, where distributional mismatch may be less harmful. Additionally, incorporating domain constraints or physiologically informed priors into generative models may improve the realism and utility of synthetic medical audio. Improvements in source data quality, including noise reduction and standardized recording protocols, may also be critical prerequisites for effective generative augmentation. In addition, future work should examine the interaction between synthetic data generation and adversarial robustness, including whether synthetic samples influence model sensitivity to noise, artifacts, or intentional perturbations in medical audio inputs \cite{dietrich2025adversarial,finlayson2019adversarial}.

This study has several limitations. First, our evaluation was conducted on a single dataset and focused on one respiratory sound classification task. While this dataset reflects a realistic moderate class imbalance, results may not generalize to other medical audio domains such as lung auscultation, heart sounds, or speech pathology. Second, we evaluated a single baseline convolutional neural network trained from random initialization. Although this design choice was intentional to isolate the contribution of synthetic data, alternative architectures or pretraining strategies may interact differently with augmented data. Third, model performance was assessed using macro-averaged F1 score and AUROC, and experiments were conducted using single training instances rather than repeated runs; as a result, formal inferential testing was not performed. Accordingly, these findings should be interpreted as preliminary and exploratory rather than as a comprehensive inferential evaluation across all possible model and augmentation combinations.

\section{Conclusion}

Our results indicate that synthetic data augmentation using current generative models does not reliably improve respiratory sound classification performance when applied directly to training data. While ensemble methods can extract limited benefit through augmentation-induced diversity, naive mixing of synthetic samples into training sets should be approached with caution. These findings underscore the importance of rigorous evaluation and suggest that, for medical audio tasks, the promise of synthetic augmentation remains conditional and highly context dependent.

\bibliographystyle{unsrt}

\begin{thebibliography}{29}

\bibitem{nanni2021ensemble}
L.~Nanni, G.~Maguolo, S.~Brahnam, and M.~Paci.
\newblock An ensemble of convolutional neural networks for audio classification.
\newblock \emph{Applied Sciences}, 2021.

\bibitem{xiao2024lungadapter}
L.~Xiao, L.~Fang, Y.~Yang, and W.~Tu.
\newblock LungAdapter: Efficient adapting audio spectrogram transformer for lung sound classification.
\newblock In \emph{Interspeech 2024}, pages 4738--4742, 2024.

\bibitem{dietrich2026evaluating}
N.~Dietrich, D.~McShannon, and M.~F. Rzepka.
\newblock Evaluating few-shot prompting for spectrogram-based lung sound classification using a multimodal language model.
\newblock \emph{PLOS Digital Health}, 5:e0001179, 2026.

\bibitem{dablain2022understanding}
D.~Dablain, K.~N. Jacobson, C.~Bellinger, M.~Roberts, and N.~Chawla.
\newblock Understanding CNN fragility when learning with imbalanced data.
\newblock \emph{arXiv preprint}, 2022.

\bibitem{yu2025advances}
S.~Yu, J.~Yu, L.~Chen, B.~Zhu, X.~Liang, Y.~Xie, et~al.
\newblock Advances and challenges in respiratory sound analysis: A technique review based on the ICBHI2017 database.
\newblock \emph{Electronics}, 14:2794, 2025.

\bibitem{rocha2021automatic}
B.~M. Rocha, D.~Pessoa, A.~Marques, P.~Carvalho, and R.~P. Paiva.
\newblock Automatic classification of adventitious respiratory sounds: A (un)solved problem?
\newblock \emph{Sensors}, 21, 2021.

\bibitem{rocha2019open}
B.~M. Rocha, D.~Filos, L.~Mendes, G.~Serbes, S.~Ulukaya, Y.~P. Kahya, et~al.
\newblock An open access database for the evaluation of respiratory sound classification algorithms.
\newblock \emph{Physiological Measurement}, 40, 2019.

\bibitem{pezoulas2024synthetic}
V.~C. Pezoulas, D.~I. Zaridis, E.~Mylona, C.~Androutsos, K.~Apostolidis, N.~S. Tachos, et~al.
\newblock Synthetic data generation methods in healthcare: A review on open-source tools and methods.
\newblock \emph{Computational and Structural Biotechnology Journal}, 23:2892--2910, 2024.

\bibitem{zhu2026audio}
G.~Zhu, Y.~Wen, and Z.~Duan.
\newblock Audio generation through score-based generative modeling: Design principles and implementation.
\newblock \emph{arXiv preprint}, 2026.

\bibitem{waseem2025review}
H.~M. Waseem, S.~U. Islam, N.~Matragkas, G.~Epiphaniou, T.~N. Arvanitis, and C.~Maple.
\newblock Review of generative AI for synthetic data generation: a healthcare perspective.
\newblock \emph{Artificial Intelligence Review}, 59:55, 2025.

\bibitem{dietrich2025greener}
N.~Dietrich and K.~Hanneman.
\newblock Greener by design: Weighing the environmental impact of radiology AI development.
\newblock \emph{Canadian Association of Radiologists Journal}, 2025.

\bibitem{gonzales2023synthetic}
A.~Gonzales, G.~Guruswamy, and S.~R. Smith.
\newblock Synthetic data in health care: A narrative review.
\newblock \emph{PLOS Digital Health}, 2:e0000082, 2023.

\bibitem{pasculli2025synthetic}
G.~Pasculli, M.~Virgolin, P.~Myles, A.~Vidovszky, C.~Fisher, E.~Biasin, et~al.
\newblock Synthetic data in healthcare and drug development: Definitions, regulatory frameworks, issues.
\newblock \emph{CPT: Pharmacometrics \& Systems Pharmacology}, 14:840--852, 2025.

\bibitem{bae2024patch}
S.~Bae, J.-W. Kim, W.-Y. Cho, H.~Baek, S.~Son, B.~Lee, et~al.
\newblock Patch-Mix contrastive learning with audio spectrogram transformer on respiratory sound classification.
\newblock \emph{arXiv preprint}, 2024.

\bibitem{kala2023constrained}
A.~Kala and M.~Elhilali.
\newblock Constrained synthetic sampling for augmentation of crackle lung sounds.
\newblock In \emph{Annual International Conference of the IEEE Engineering in Medicine and Biology Society}, pages 1--5, 2023.

\bibitem{ghosh2025synthio}
S.~Ghosh, S.~Kumar, Z.~Kong, R.~Valle, B.~Catanzaro, and D.~Manocha.
\newblock Synthio: Augmenting small-scale audio classification datasets with synthetic data.
\newblock \emph{arXiv preprint}, 2025.

\bibitem{bhattacharya2023coswara}
D.~Bhattacharya, N.~K. Sharma, D.~Dutta, S.~R. Chetupalli, P.~Mote, S.~Ganapathy, et~al.
\newblock Coswara: A respiratory sounds and symptoms dataset for remote screening of SARS-CoV-2 infection.
\newblock \emph{Scientific Data}, 10:397, 2023.

\bibitem{ronneberger2015unet}
O.~Ronneberger, P.~Fischer, and T.~Brox.
\newblock U-Net: Convolutional networks for biomedical image segmentation.
\newblock \emph{arXiv preprint}, 2015.

\bibitem{prodhan2008wheeze}
P.~Prodhan, R.~S. Dela~Rosa, M.~Shubina, K.~E. Haver, B.~D. Matthews, S.~Buck, et~al.
\newblock Wheeze detection in the pediatric intensive care unit: Comparison among physician, nurses, respiratory therapists, and a computerized respiratory sound monitor.
\newblock \emph{Respiratory Care}, 53, 2008.

\bibitem{kim2021respiratory}
Y.~Kim, Y.~Hyon, S.~S. Jung, S.~Lee, G.~Yoo, C.~Chung, et~al.
\newblock Respiratory sound classification for crackles, wheezes, and rhonchi in the clinical field using deep learning.
\newblock \emph{Scientific Reports}, 11:17186, 2021.

\bibitem{sarkar2015auscultation}
M.~Sarkar, I.~Madabhavi, N.~Niranjan, and M.~Dogra.
\newblock Auscultation of the respiratory system.
\newblock \emph{Annals of Thoracic Medicine}, 10:158, 2015.

\bibitem{chu2023cnn}
H.-C. Chu, Y.-L. Zhang, and H.-C. Chiang.
\newblock A CNN sound classification mechanism using data augmentation.
\newblock \emph{Sensors}, 23, 2023.

\bibitem{karim2025ensemble}
A.~Karim, S.~Ryu, and I.~C. Jeong.
\newblock Ensemble learning for biomedical signal classification: a high-accuracy framework using spectrograms from percussion and palpation.
\newblock \emph{Scientific Reports}, 15:21592, 2025.

\bibitem{nadkarni2024afen}
R.~Nadkarni, E.~Nikolakakis, and R.~Marinescu.
\newblock AFEN: Respiratory disease classification using ensemble learning.
\newblock \emph{arXiv preprint}, 2024.

\bibitem{dietrich2025agentic}
N.~Dietrich.
\newblock Agentic AI in radiology: emerging potential and unresolved challenges.
\newblock \emph{British Journal of Radiology}, 2025.

\bibitem{qureshi2026improving}
A.~M. Qureshi, A.~Kaushik, R.~Loughran, K.~McDaid, and F.~McCaffery.
\newblock Improving medical data quality via synthetic data generation: a review.
\newblock \emph{Network Modeling Analysis in Health Informatics and Bioinformatics}, 15:46, 2026.

\bibitem{chakraborty2025generative}
S.~Chakraborty, P.~Kochhar, S.~Patil, K.~Kotecha, S.~Gite, G.~Selvachandran, et~al.
\newblock Generative adversarial network augmented data for improved heart sound abnormality detection.
\newblock \emph{Computers in Biology and Medicine}, 195:110623, 2025.

\bibitem{dietrich2025adversarial}
N.~Dietrich, B.~Gong, and M.~N. Patlas.
\newblock Adversarial artificial intelligence in radiology: Attacks, defenses, and future considerations.
\newblock \emph{Diagnostic and Interventional Imaging}, 2025.

\bibitem{finlayson2019adversarial}
S.~G. Finlayson, J.~D. Bowers, J.~Ito, J.~L. Zittrain, A.~L. Beam, and I.~S. Kohane.
\newblock Adversarial attacks on medical machine learning.
\newblock \emph{Science}, 363:1287--1289, 2019.

\end{thebibliography}

\end{document}